%% file: BinaryTree160513.tex
\documentclass[12pt]{iopart}
\usepackage{iopams}

\expandafter\let\csname equation*\endcsname\relax 
\expandafter\let\csname endequation*\endcsname\relax

\usepackage{amsmath}
\usepackage{mathrsfs}
\usepackage[dvipdfmx]{graphicx}

\usepackage{citesort}

\usepackage[dvips]{color}

\definecolor{darkgreen}{rgb}{0.0, 0.5, 0.0}
\definecolor{purple}{rgb}{0.50, 0.0, 0.50}

\newcommand{\blue}[1]{\textcolor{blue}{#1}}
\def\be{\begin{eqnarray}}
\def\ee{\end{eqnarray}}

\def\mrm{\mathrm}

\def\vphi{\varphi}
\def\toinf{\to \infty}

\def\To{\Rightarrow}
\def\uto{\nearrow}
\def\dto{\searrow}

\def\TT{\hat{T}}
\def\VV{\hat{V}}
\def\EE{\hat{E}}
\def\CC{\hat{C}}

\def\PP{\hat{P}}
\def\QQ{\hat{Q}}

\def\oo{{\hat{o}}}
\def\qq{\hat{q}}
\def\cchi{\hat{\chi}}

\def\ppsi{\hat{\psi}}
\def\qqq{\tilde{q}}
\def\qqqq{\tilde{\tilde{q}}}

\def\sums{\sum_{s=1}^\infty}
\def\sumt{\sum_{t=1}^\infty}

\def\pb{\overline{p}}
\def\mm{{(m)}}

\begin{document}

\title
% [Local cluster-size statistics in the critical phase of bond percolation on the Cayley tree ]
{Local cluster-size statistics in the critical phase of bond percolation on the Cayley tree}

\author{Tomoaki Nogawa$^1$, Takehisa Hasegawa$^2$ and Koji Nemoto$^3$}

\address{$^1$ Faculty of Medicine, Toho University, 5-21-16, Omori-Nishi, Ota-ku, Tokyo, 143-8540, Japan}
\address{$^2$ Department of Mathematics and Informatics, Ibaraki University, 2-1-1, Bunkyo, Mito, 310-8512, Japan}
\address{$^3$ Department of Physics, Hokkaido University, Kita 10 Nishi 8, Kita-ku, Sapporo, Hokkaido, 060-0810, Japan}

\ead{nogawa@med.toho-u.ac.jp, 
takehisa.hasegawa.sci@vc.ibaraki.ac.jp and 
nemoto@statphys.sci.hokudai.ac.jp}

\vspace{10pt}
\begin{indented}
\item[]February 2014
\end{indented}

\begin{abstract}
We study bond percolation of the Cayley tree (CT) by focusing on 
the probability distribution function (PDF) of a local variable, 
namely, the size of the cluster including a selected vertex. 
Because the CT does not have a dominant bulk region, which is free from the boundary effect, 
even in the large-size limit, the phase of the system on it is not well defined. 
We herein show that local observation is useful to define the phase of such a system 
in association with the well-defined phase of the system on the Bethe lattice, 
that is, an infinite regular tree without boundary. 
Above the percolation threshold, the PDFs of the vertex at the center of the CT (the origin) 
and of the vertices near the boundary of the CT (the leaves) have different forms, 
which are also dissimilar to the PDF observed in the ordinary percolating phase of a Euclidean lattice. 
The PDF for the origin of the CT is bimodal: 
a decaying exponential function and a system-size-dependent asymmetric peak, 
which obeys a finite-size-scaling law with a fractal exponent. 
These modes are respectively related to the PDFs of the finite and infinite clusters 
in the nonuniqueness phase of the Bethe lattice.
On the other hand, the PDF for the leaf of the CT is a decaying power function. 
This is similar to the PDF observed at a critical point of a Euclidean lattice 
but is attributed to the nesting structure of the CT around the boundary. 
\end{abstract}

% \vspace{2pc}
\noindent{\it Keywords}: Classical phase transitions (Theory), Percolation problems (Theory), Random graphs, networks
% XXXXXX, YYYYYYYY, ZZZZZZZZZ
%\maketitle

\section{Introduction}

In recent years, statistical physics on complex networks \cite{Newman03} has been intensively developed \cite{Dorogovtsev08}. 
Some collective phenomena on complex networks, such as percolation and the Ising model, 
show specific behaviors that are not observed in the systems on Euclidean lattices. 
For instance, a system exhibits behaviors similar to those observed in a system at a critical {\it point}, 
such as divergence of the susceptibility, in a {\it finite range} of the model parameter \cite{Hasegawa-Nogawa-Nemoto14}, 
which we call a critical {\it phase}.
This type of critical region has been observed in percolation and the Ising model on Cayley trees (CTs) \cite{Hasegawa-Nogawa-Nemoto14}, hierarchical small-world networks \cite{Hasegawa-Sato2010,Boettcher12}, and growing random networks \cite{Hasegawa10c}. 
On the basis of the renormalization group analysis of the systems on hierarchical small-world networks \cite{Boettcher-Brunson11,Nogawa-Hasegawa14}, 
we speculate that small-world, i.e., infinite dimensional, and hierarchical properties of graphs play essential roles.

On inhomogeneous graphs in which vertices are not transitive, 
it is not trivial whether a phase of systems with large degrees of freedom 
can be defined by using macroscopic state variables 
in the same manner for systems on homogeneous graphs, e.g., Euclidean lattices. 
In the first place, we say that a system is in a single phase 
when the property of the system is homogeneous; 
if we divide a system into multiple macroscopic pieces of the same size, 
all pieces take the same value of extensive variables such as magnetization. 
In equilibrium, we suppose that the macroscopic state and the phase 
of a system are uniquely identified by a small number of extensive variables 
that is far fewer than the total degrees of freedom. 
It is natural to consider that the feasibility of the reduction in degrees of freedom is to a large extent dependent on the homogeneity. 
On the other hand, general inhomogeneous graphs cannot be divided into equal pieces, even in the macroscopic scale. 
Therefore we need further information other than extensive variables 
to describe systems on an inhomogeneous graph. 
On hierarchical graphs, small number of vertices often have a dominant influence on the state of the system. 
Furthermore local variables on such vertices probably 
behave quite differently from extensive variables, 
which are usually defined as summations of local variables without weighting.
To adequately characterize the state of an inhomogeneous system, 
% A candidate solution to characterize the state of the system adequately 
it is useful to observe local variables at representative vertices.

In this paper, we consider percolation on a CT 
to study the statistics of local variables. 
A CT is a subgraph of a Bethe lattice (BL), which is an infinite regular graph without cyclic path; 
an $n$-layer CT is an $(n-1)$-ball centered at a vertex of a BL. 
The boundary of a CT is given by leaves: 
the vertices with degree 1 in the outermost layer. 
The number fraction of the leaves remains positive for $n \toinf$, 
and therefore a CT does not have a dominant bulk region free from the boundary effect 
even in the limit $n \toinf$ \cite{Ostilli12}. 
This is in contrast with the finite $d$-dimensional Euclidean lattice, 
where the boundary is $(d-1)$-dimensional and its fraction goes to zero in the large-size limit.  
The existence of the boundary brings inhomogeneity to a CT. 
When we consider the shortest paths between vertices, 
a few vertices near the origin are frequently used. 
Therefore the betweenness centrality is smaller for vertices in more outer layers; 
more important vertices have smaller populations. 
Inhomogeneity has an influence on the type of phase transition; 
whereas the Ising model on a BL undergoes a second-order transition \cite{Kurata53,Domb60,Fisher61}, 
that on a CT undergoes an infinite-order transition \cite{Eggarter74,Muller-Hartmann74,Matsuda74}. 
Similar behavior is observed in the bond-percolation transitions on a CT and a BL \cite{Hasegawa-Nogawa-Nemoto14}, 
which will be reported in detail in the following sections. 
% \green{We should note that ...}
We also note that the continuous transition on a BL is known to be similar to those in the mean-field universality class, 
but its supercritical phase is not an ordinary percolating phase; 
an infinite cluster is not unique, but the number of infinite clusters is infinite 
\cite{Grimmet-Newman90,Benjamini-Schramm96}. 
Such a phase is called a nonuniqueness phase \cite{Lyons00}. 
For percolation on a CT, the same parameter region is a critical phase \cite{Hasegawa-Nogawa-Nemoto14}, 
where the cluster size histogram exhibits a power-law distribution. 
This paper aims to understand the critical phase of an inhomogeneous graph, a CT,  
in association with the nonuniqueness phase of a homogeneous graph, a BL. 
For this purpose, we investigate the probability distribution function (PDF) of the size of a locally-defined cluster 
and its moment for percolation on a CT. 
We show that the PDF of the vertices in the outer layers in the supercritical phase 
is similar to that observed at critical points of homogeneous systems. 
On the other hand, the PDFs of the vertices near the origin 
converge to that for a BL in the large-size limit, 
but an unignorable finite-size effect, which is related to an infinite number of infinite clusters, 
causes another kind of singular behavior for the moments. 
We propose two distinct finite-size-scaling laws for PDFs with respect to the two regimes.

%%%%%%%%%%%%%%%%%%%%%%%%%%%%%%%%%%%%%%%%%%%%%%%%%%%%%%%%%%%%%%%%%%
\section{Preliminary}
\label{sub:preliminary}

%%%%%%%%%%%%%%%%%%%%%%%%%%%%%%%%%%%%%%%%%%%%%%%%%%%%%%%%%%%%%%%%%
\begin{figure}[t]
% \hspace{0.1cm}{\bf (a)}\hspace{7.05cm}{\bf (b)}\\ \vspace{-1.2cm}
\begin{center}
% \hspace{-6.8cm} {\bf{\large (a)}}\\ \vspace{-14.1pt}
\includegraphics[trim=170 158 290 60,scale=0.35,clip]{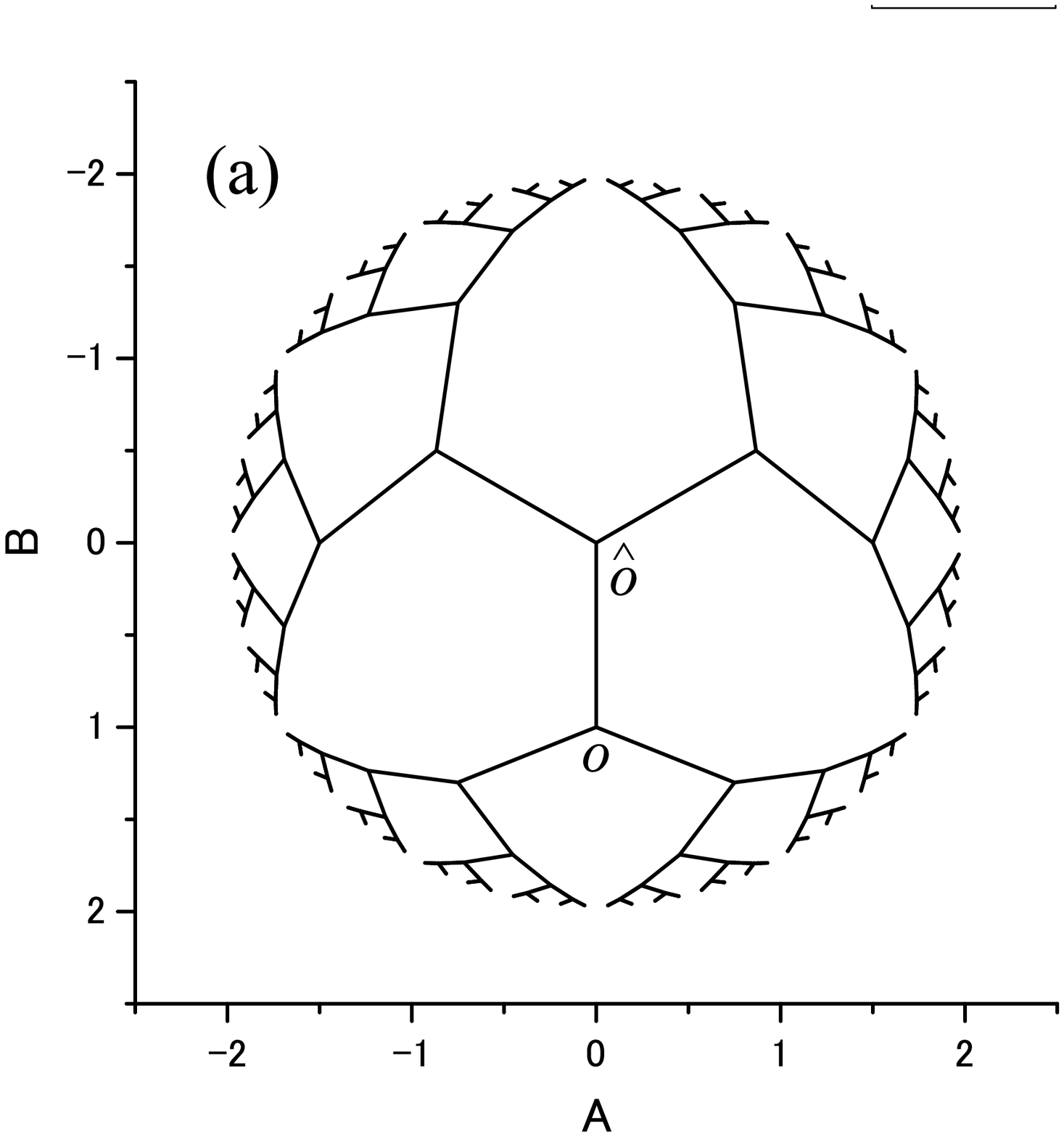}
\hspace{1cm}
\includegraphics[trim=170 125 210 70,scale=0.35,clip]{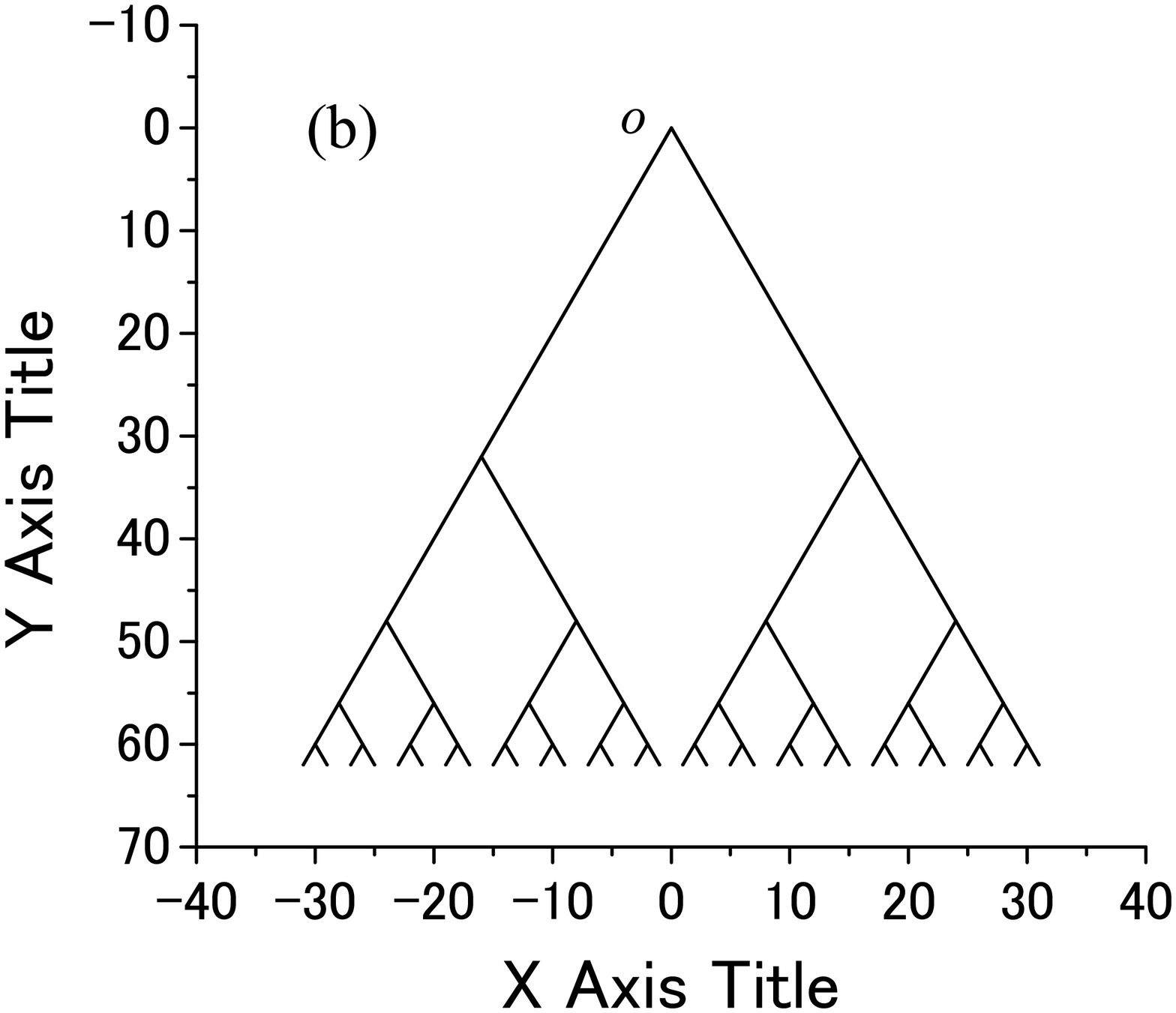}
\end{center}
\vspace{-5mm}
\caption{\label{fig:tree}
(a) Cayley tree $\TT_7$ (b) binary tree $T_6$. 
}\label{fig:tree}
\end{figure}
%%%%%%%%%%%%%%%%%%%%%%%%%%%%%%%%%%%%%%%%%%%%%%%%%%%%%%%%%%%%%%%%%

\subsection{trees}

Let $\TT=(\VV, \EE)$ be the BL where the degree of every vertex is three. 
Here $\VV$ and $\EE$ denote the set of vertices and the set of edges, respectively. 
We arbitrarily choose a vertex $\oo \in \VV$ as the origin of $\TT$. 

Let $\TT_n = (\VV_n, \EE_n)$ be the CT with $n$ layers, that is an $(n-1)$-ball in $\TT$ centered at $\oo$. 
Figure~\ref{fig:tree}(a) shows $\TT_7$. 
We say that $v \in \TT_n$ is in the $l$th layer when $d(\oo,v)$: the shortest path length between $\oo$ and $v$ 
equals $l \in \{0, 1, \cdots, n-1\}$. 
We call the vertices in the $(n-1)$th layer leaves, whose degrees are unity. 
The numbers of the vertices and the edges are $|\VV_n|=3 \times 2^{n-1}-2$ and $|\EE_n|=|\VV_n|-1$, respectively. 

By removing $\oo$ and three edges connecting to it from $\TT$, we obtain three infinite binary trees. 
Let $T=(V, E)$ denote one of them. 
We denote by $o$ the vertex that once connected to $\oo$ and call it the root. 
Let $T_n = (V_n, E_n)$ be the binary tree with $n$ layers, that is an $(n-1)$-ball in $T$ centered at $o$. 
Figure~\ref{fig:tree}(b) shows $T_6$. 
We define the layers of $T_n$ similarly for $\TT_n$, and then the vertices in the $(n-1)$th layers are leaves. 
% $\TT_{n+1}$ is made by preparing three copies of $T_n$ and an isolated vertex  
% and giving an edge between the isolated vertex and each root of $T_n$'s. 
The numbers of the vertices and the edges are $|V_n|=2^n-1$ and $|E_n|=|V_n|-1$, respectively. 
Hereafter, we assume $|V_n| \gg 1$, and then approximate $|V_n|$ by $2^n$.

\subsection{percolation}

Let us consider bond percolation for $T_n$. 
Each $e \in E_n$ is independently open with a unique probability $p$ and closed with probability $1 - p \equiv \pb$.  
The connected components of the open edges are called clusters. 
We denote the set of vertices in the cluster including $v \in V_n$ by $C_{nv}(\subseteq V_n)$ 
and denote the cluster size, that is the number of vertices in $C_{nv}$, by $|C_{nv}|$. 
We also consider bond percolation for $\TT_n$, $T$, and $\TT$. 
Then, we similarly define $\CC_{nv}$ for $\TT_n$, $C_v$ for $T$ and $\CC_{v}$ for $\TT$.

\subsection{probability distribution functions (PDFs) of cluster sizes}

Here we introduce some PDFs of cluster sizes. 
For $v \in T_n$, we define 
\be
q_{nv}(s) \equiv P(|C_{nv}|=s) = \left\langle \delta_{s |C_{nv}|} \right\rangle. 
\ee
Hereafter, $P(X)$ is the probability of a proposition $X$, which is a function of $p$ (but we omit to note) 
and $\langle x \rangle$ is the expectation value of $x$. 
Similarly, we define $\qq_{nv}(s) \equiv P(|\CC_{nv}|=s)$ for $\TT_n$.  
For $\TT$,  we define a unique PDF $\qq(s) \equiv P(|C_\oo|=s)$ because $P(|C_v|=s)$ does not depend on $v$. 
Similarly, we define $q(s) \equiv P(|C_o|=s)$ for $T$. 

Remembering that $T_n \subset T$, we have 
\be
q_{n o}(s) 
&=& P\left( |C_o \cap V_n| = s \right) 
\nonumber \\
&=& P\left( C_o \subseteq V_n \wedge |C_o|=s \right)
+ P\left( C_o \not\subseteq V_n \wedge |C_o \cap V_n| = s \right). 
% \\
% &=& q(s) + \sum_{s'=n+1}^\infty P\left( \CC \cap (\VV - \VV_n) \ne \phi 
% \wedge |\CC| = s' \wedge |\CC \cap \VV_n| = s \right)
\label{eq:P_divide}
\ee
From the fact that 
$|C_o \cap V_n| \le n-1 \To C_o \subset V_n$
%\green{$s \le n$でなく$s \le n-1$でなければいけないのは、$C_o$ではなく$|C_o \subset V_n|$に対する条件を考えた結果として得られます。$|V_n|$からはみ出す条件$C_o \not\subseteq V_n \To |C_o \cap V_n| \ge n$の対偶をとると青字の式が出ます。対偶をとる前の式のほうがわかりやすいでしょうか？}
, we have 
\be
q_{no}(s) = q(s) \ \  \mrm{for} \ \  s \le n-1
\quad 
\To \quad
\lim_{n \toinf} q_{n o}(s) = q(s) \ \ \mrm{for} \ \ s < \infty. 
%  \quad \mrm{for} \quad  s<\infty. 
\label{eq:limit_of_q_n}
\ee
In a similar way, we have 
$\lim_{n \toinf} \qq_{n \oo}(s) = \qq(s)$. 
Furthermore, we have 
$\lim_{n \toinf} \qq_{n v}(s) = \qq(s)$ if $d(\oo,v) < \infty$. 
Note that similar relation does not hold for $q_{nv}$ unless $v=o$.

\subsection{local fractal exponent}

To quantify the upper cutoff of the PDF, 
we define a local fractal exponent for $v \in V_n$ as 
\be
\psi_v \equiv \inf 
\left\{ \vphi \ {\big |}\  \vphi > 0, \ 
\lim_{n \toinf} P\left( |C_{nv}|>|V_n|^{\vphi} \right) = 0 \right \}. 
\label{eq:def:vphi}
\ee
Similarly we define $\ppsi_v$ for $v \in \VV_n$. 
% This measure a kind of the order of the diverging clusters at $v$. 
For percolation on an Euclidean lattice, 
this exponent equals zero in a nonpercolating phase 
and unity in a percolating phase. 
In the former, the PDF of a cluster size decays exponentially. 
In the latter, the PDF has a Gaussian peak 
whose position is asymptotically proportional to the number of total vertices. 
This peak is related to a unique giant component for each sample. 
% $\psi  = \lim_{n \toinf} \frac{\ln \chi_{no}}{\ln |V_n|}, $を一般に示せないか？

\subsection{moments}

We define the $m$-th order moment of $q_{nv}(s)$ for $T_n$ as 
\be
\chi_{nv}^\mm \equiv \langle |C_{nv}|^m \rangle = \sums s^m q_{nv}(s).  
\ee
Similarly, we define 
$\cchi_{nv}^\mm$ for $\qq_{nv}(s)$, 
$\chi^\mm$ for $q(s)$ and 
$\cchi^\mm$ for $\qq(s)$. 
Note that $\chi^\mm$ and $\cchi^\mm$ does not include the contribution from infinite clusters 
and therefore does not necessarily diverge 
even if an infinite cluster exists with positive probability. 
% These moments correspond to the susceptibility to a local magnetic field of the Potts model. 

\subsection{Generating functions (GFs)}

We define the GF for $q_{nv}$ as 
\be
Q_{nv}(x) \equiv \sum_{s=1}^{\infty} q_{nv}(s) x^s.
\label{eq:Q_n(x)}
\ee
% Again, we note that $\QQ(x)$ does not include the contribution from the infinite clusters. 
The moments of a PDF is calculated by differentiating the corresponding GF, e.g., 
\be
\chi_{nv}^\mm = Q_{nv}^\mm(1) \equiv \left. \left( x \frac{d}{dx} \right)^{\!\! m} \! Q_{nv}(x) \right|_{x=1} .
\label{eq:Q_n(x)}
\ee
Similarly, we define 
$\QQ_{nv}(x)$ for $\qq_{nv}(s)$, 
$Q(x)$ for $q(s)$ and 
$\QQ(x)$ for $\qq(s)$.

From the nesting structure of $T_n$, 
we have a recursion equation for $Q_{no}(x)$ as 
\be
Q_{n+1,o}(x) = x [ p Q_{no}(x) + \pb ]^2. 
\label{eq:recursion_Q}
\ee
With $Q_{no}(x)$, the recursion equation for $Q_{no}(x)$ is given by 
\be
\QQ_{n+1,o}(x) = x [ p Q_{no}(x) + \pb ]^3 = Q_{n+1,o}(x) [ p Q_{no}(x) + \pb ]. 
\label{eq:recursion_QQ}
\ee

%%%%%%%%%%%%%%%%%%%%%%%%%%%%%%%%%%%%%%%%%%%%%%%%%%%%%%%%%%%%%%%%%%%%%%%%%%%%%%%%%%%%%%%%%%%%%%%%
\section{Results}
\label{sec:results}

In this section, we investigate the PDF and the singularities of the moments, 
which are related to the analyticity of GFs as a function of $p$ and $x$. 
% to determine the singular behavior of percolation. 
We focus on the results for $T_n$ and $T$,  
but their properties are almost the same with those for $\TT_n$ and $\TT$, respectively.

\subsection{Infinite trees}

First we overview the results for the infinite binary tree, $T$ 
\cite{Flory41-2,Ziff80}. 
From Eq.~\eqref{eq:limit_of_q_n}, we have $Q(x) = \lim_{n \toinf} Q_n(x)$. 
By evaluating the fixed point of Eq.~\eqref{eq:recursion_Q}, we obtain 
\be
Q(x) = \frac{ 1 - 2p \pb x - \sqrt{ 1 - 4 p \pb x } }{ 2 p^2 x }, 
\quad
Q^{(1)}(x) = \frac{ x [\pb + p Q(x)]^2 }{1 - 2 p x [ \pb + p Q(x) ]},  
\label{eq:Q_inf(x)}
\\
q(s) \propto s^{-3/2} (4 p \pb)^s \quad\mrm{for}\quad s \gg 1. 
\label{eq:fixed_point}
\ee
See \ref{sec:fixed_point} and \ref{sec:series_expansion} for the derivation. 
The transition threshold is given by $p=p_c=1/2$, at which $4 p \pb$ equals unity 
and then the exponential cutoff of $q(s)$ diverges.

From the second equation of Eq.~\eqref{eq:Q_inf(x)}, 
the probability that $o$ is included in a finite cluster is given by 
\be
Q(1) = \sums q(s) = \frac{ 2 p^2 + (1 - 2p) - | 1 - 2p | }{2p^2} 
= \left \{
\begin{array}{ccc}
1 & \mrm{for} & p \le p_c
\\
(\pb/p)^2 & \mrm{for} & p > p_c
\end{array}
\right. 
.
\label{eq:chi_inf(p)}
\ee
Thus, the probability that $o$ is included in an infinite cluster is given by 
\be
P_\infty \equiv 1 - Q(1) 
= \left \{
\begin{array}{ccc}
0 & \mrm{for} & p \le p_c
\\
1 - (\pb/p)^2 & \mrm{for} & p > p_c
\end{array}
\right. 
.
\ee
For $p \dto p_c$, we have a power law 
\be
P_\infty \propto (p-p_c)^\beta, \quad \beta=1.
\label{eq:cri_0_bl}
\ee

From Eq.~\eqref{eq:Q_inf(x)} and Eq.~\eqref{eq:chi_inf(p)}, we have 
\be
\chi^{(1)} = Q^{(1)}(1) = \left \{
\begin{array}{ccc}
\displaystyle
\frac{1}{1-2p} & \mrm{for} & p < p_c
\\
\displaystyle
\left(\frac{\pb}{p}\right)^2  \frac{1}{2p-1} & \mrm{for} & p > p_c
\end{array}
\right. 
.
\ee
In the limits $p \uto p_c$ and $p \dto p_c$, $\chi^{(1)}$ diverges 
and a power-law 
\be
\chi^{(1)} \propto |p-p_c|^{-\gamma}, \quad \gamma=1, 
\label{eq:cri_1_bl}
\ee
holds. 
The higher moment, $\chi^\mm$ with $m>1$, is also finite for $p \ne p_c$ 
because $q(s)$ decays exponentially with increasing $s$ for $p \ne p_c$.

For $\TT_n$, we have 
\be
\PP_\infty \equiv 1 - \QQ(1) = \left \{
\begin{array}{ccc}
0 & \mrm{for} & p \le p_c
\\
1 - (\pb/p)^3 & \mrm{for} & p > p_c
\end{array}
\right. 
,
\nonumber \\ 
\cchi^{(1)} = \left \{
\begin{array}{ccc}
\displaystyle
\frac{1+p}{1-2p} & \mrm{for} & p < p_c
\\
\displaystyle
\left(\frac{\pb}{p}\right)^3  \frac{2-p}{2p-1} & \mrm{for} & p > p_c
\end{array}
\right. 
.
\ee
Power-laws same as Eq.~\eqref{eq:cri_0_bl} and Eq.~\eqref{eq:cri_1_bl} 
hold for $\PP_\infty$ and $\cchi^{(1)}$, respectively.

%%%%%%%%%%%%%%%%%%%%%%%%%%%%%%%%%%%%%%%%%%%%%%%%%%%%%%%%%%%%%%%%%%%%%%%%%%%%%%%

%%%%%%%%%%%%%%%%%%%%%%%%%%%%%%%%%%%%%%%%%%%%%%%%%%%%%%%%%%%%%%%%%%%%%%%%%%%%%%%
\subsection{Root}

Here we consider the PDF for the cluster including the root of the finite binary tree.

\subsubsection{PDF}

%%%%%%%%%%%%%%%%%%%%%%%%%%%%%%%%%%%%%%%%%%%%%%%%%%%%%%%%%%%%%%%%%
\begin{figure}[t]
% \hspace{0.1cm}{\bf (a)}\hspace{7.05cm}{\bf (b)}\\ \vspace{-1.2cm}
\begin{center}
% \hspace{-6.8cm} {\bf{\large (a)}}\\ \vspace{-14.1pt}
\includegraphics[trim=20 20 170 10,scale=0.33,clip]{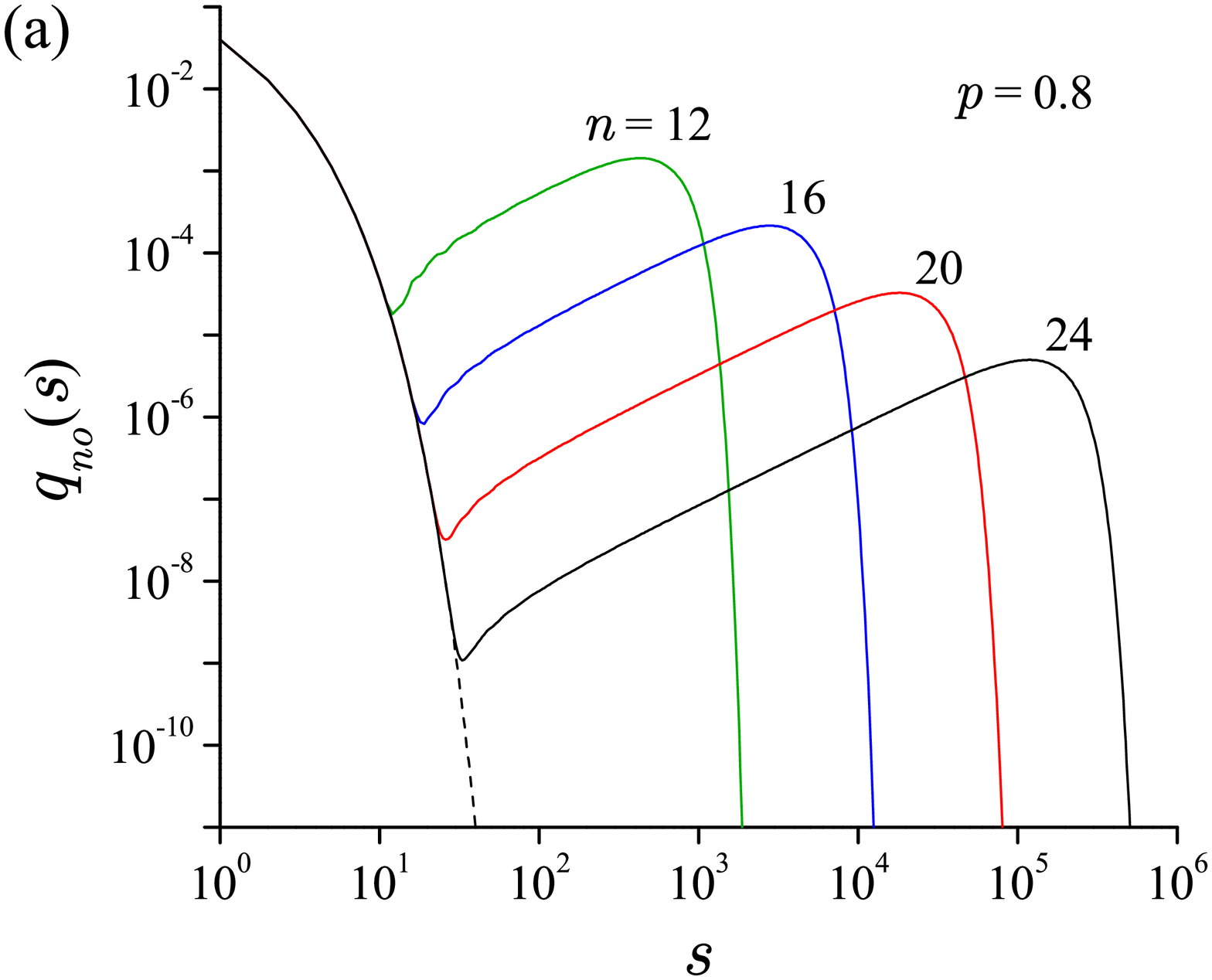}
\includegraphics[trim=20 20 170 10,scale=0.33,clip]{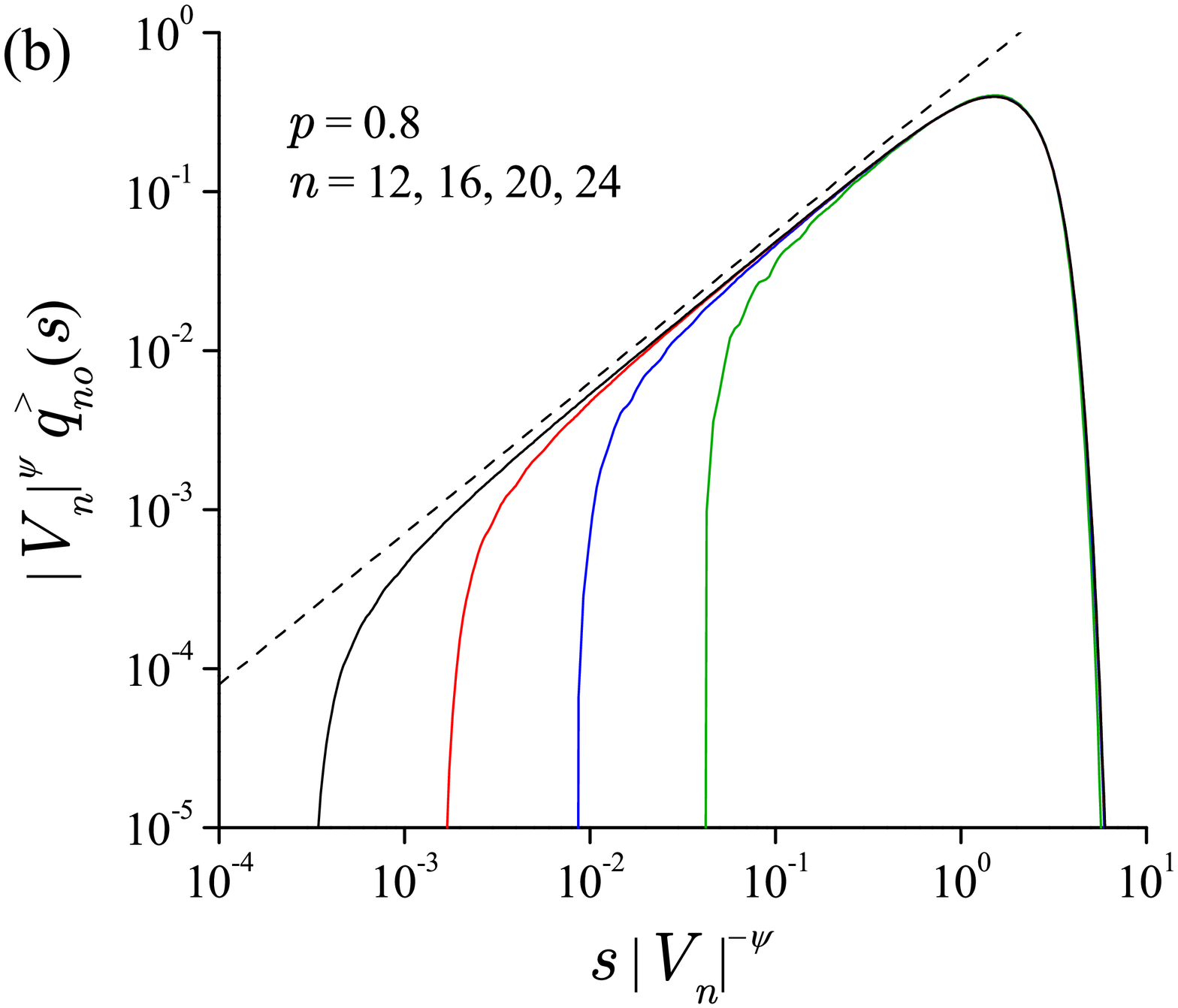}
\end{center}
\vspace{-5mm}
\caption{\label{fig:rn-s}
(a) Numerical results of $q_{no}(s)$ for $p=0.8>p_c$ and several $n$'s, 
which is calculated by Eq.~\eqref{eq:term_recursion}. 
The dashed line indicates $q(s)$. 
(b) scaling result of $q_{no}^>(s)$. 
The slope of the dashed line is $\tau'$.
}
\end{figure}
%%%%%%%%%%%%%%%%%%%%%%%%%%%%%%%%%%%%%%%%%%%%%%%%%%%%%%%%%%%%%%%%%

As mentioned before, $q_{no}(s)$ converges to $q(s)$ for $n \toinf$. 
For $p<p_c$, $q_{no}(s)$ is very similar to $q(s)$ except the existence of the finite size cutoff. 
For $p>p_c$, $q_{no}(s)$ has characteristic bimodal structure 
as indicated in Fig.~\ref{fig:rn-s}(a). 
The separation of the two modes becomes clearer as $n$ increases. 
The small-$s$ part converges to $q(s)$ for $n \to \infty$. 
On the other hand, the large-$s$ part forms an asymmetric peak, 
whose position moves to the larger $s$ direction 
and the height decreases toward zero with increasing $n$. 
Let us denote the large-$s$ part by 
\be
q^>_{no}(s) \equiv q_{no}(s) - q(s).  
\ee
Whereas $q(s)$ corresponds to the part of finite clusters for the PDF in $T$, 
$q^>_{no}(s)$ is related to the statistical property of an infinite cluster in $T$; 
for large $s$ such that $q(s) \ll q_{no}(s)$, 
$q_{no}(s)$ corresponds to the second term on the r.h.s. of Eq.~\eqref{eq:P_divide} 
and we have 
\be
q^>_{no}(s) \approx P\left( |C_o|=\infty \wedge |C_o \cap V_n|=s \right). 
\ee
We remark that $\sums q^>_{no}(s) = P_\infty$ and $q^>_{no}(s) = 0$ for $s \le n-1$.

Figure~\ref{fig:rn-s}(a) indicates that the shape of $q^>_{no}(s)$'s 
for various $n$ are congruent in the double-logarithmic plot. 
We assume that $q^>_{no}(s)$ asymptotically obeys a power-law: 
\be
q^>_{no}(s) = |V_n|^{-\psi} \qqq^>( s |V_n|^{-\psi} )
= s^{-1} \tilde{\qqq}^>( s |V_n|^{-\psi} ) 
\label{eq:scaling_q_{no}}
\ee
for $s \gg 1$. 
Here $\qqq^>(x)$ is a scaling function and $\tilde{\qqq}^>(x) \equiv x \qqq^>(x)$. 
As to the scaling exponent $\psi$, we will show 
\be
\psi = \log_2 (2p) 
\label{eq:psi}
\ee
in Sec.~\ref{sec:psi}. 
The exponent $\psi$ monotonically increases from 0 to 1 as $p$ increases from $p_c$ to 1. 
Numerical result in Fig.~\ref{fig:rn-s}(b) supports our scaling assumption. 
Moreover, the scaling function is a power function as 
\be
\qqq^>(x) \propto x^{\tau'} \ \ \mrm{for} \ \ x \ll 1, 
\quad
\tau' = - \frac{\ln 2 \pb}{\ln 2 p} - 1 . 
\ee
(The evaluation of $\tau'$ is shown in \ref{sec:tau'}.) 
The exponent $\tau'$ equals zero at $p=p_c$ 
and monotonically increases with $p$ to diverge for $p \to 1$.

\subsubsection{local fractal exponent}

By using Eq.~\eqref{eq:scaling_q_{no}}, we have 
\be
P(|C_{no}|>|V_n|^{\vphi}) \approx 
\int_{|V_n|^{\vphi}}^\infty ds |V_n|^{-\psi} \qqq^> (s |V_n|^{-\psi})
= \int_{|V_n|^{\vphi-\psi}}^\infty dx \qqq^>(x).   
% \xrightarrow[n \toinf]{} 0
\ee
In the first equation, we omit the contribution from $q(s)$. 
In the limit $n \toinf$, 
this goes to zero for $\vphi > \psi$ 
and converges to a positive constant, that is, $P_\infty$ for $\vphi < \psi$. 
Thus, we have 
\be
\psi_o = \psi. 
\ee

\subsubsection{moments}
\label{sec:psi}

Next, we consider the asymptotic behavior of $\chi_{no}^\mm$ for $p>p_c$ and $n \toinf$. 
From Eq.~\eqref{eq:scaling_q_{no}}, we have 
\be
\chi_{no}^\mm 
& \approx & \chi^\mm + \int_1^\infty ds s^m |V_n|^{-\psi} \qqq^>(s |V_n|^{-\psi})
\nonumber \\
& \approx & |V_n|^{m \psi} \int_{|V_n|^{-\psi}}^\infty dx x^m \qqq^>(x) 
\propto  |V_n|^{m \psi} 
\label{eq:chi_rt_dis}
\ee
for $n \toinf$. 
In the second approximation,  we ignored $\chi^\mm$, which is finite for $n \to \infty$. 
Consequently, $\chi_{no}^\mm$ with arbitrary $m>0$ diverges in the limit $n \toinf$ for $p > p_c$.

For $m=1$, we can exactly calculate $\chi_{no}^{(1)} = \langle |C_{no}| \rangle$ 
and estimate $\psi$
by comparing its asymptotic form and Eq.~\eqref{eq:chi_rt_dis}.  
We have $P(v \in C_{no}) = p^{d(o,v)}$, which leads to 
\be
\chi_{no}^{(1)} = \left \langle \sum_{v \in C_{no}} 1 \right \rangle 
= \sum_{v \in V_n} P(v \in C_{no}) 
% = \sum_{v \in V_n} p^{d(o,v)} 
= \sum_{l=0}^{n-1} 2^l p^l 
= \frac{ (2p)^n - 1}{2p -1}
\ee
for $p \ne p_c$. 
(At $p=p_c$, $\chi_{no}^{(1)}$ equals $n \approx \log_2 |V_n|$.)
For $p > p_c$ and $n \toinf$, $\chi_{no}^{(1)}$ is proportional to 
$(2p)^n \propto |V_n|^{\log_2(2p)}$. 
Thus we have Eq.~\eqref{eq:psi}.
% \be
% \psi = \log_2(2p).
% \label{eq:psi}
% \ee 

%%%%%%%%%%%%%%%%%%%%%%%%%%%%%%%%%%%%%%%%%%%%%%%%%%%%%%%%%%%%%%%%%%%%%%%%%%%%%%%%%%%%%%%%%%%%%%%%%%
\subsection{Leaf}

Let $v_l$ be a vertex distant from the outermost layer by $l$, 
i.e., in the $(n-l-1)$th layer of $T_n$. 
% (see Fig.~\ref{fig:tree2})
In this subsection, we consider the limit $n \toinf$ with fixing $l$. 
% , where $v_l \in V_n$ is identified with $v_{l+1} \in V_{n+1}$. 

\subsubsection{PDF}
\label{sec:pdf4leaf}
%%%%%%%%%%%%%%%%%%%%%%%%%%%%%%%%%%%%%%%%%%%%%%%%%%%%%%%%%%%%%%%%%
\begin{figure}[t]
% \hspace{0.1cm}{\bf (a)}\hspace{7.05cm}{\bf (b)}\\ \vspace{-1.2cm}
\begin{center}
% \hspace{-6.8cm} {\bf{\large (a)}}\\ \vspace{-14.1pt}
\includegraphics[trim=20 40 170 10,scale=0.33,clip]{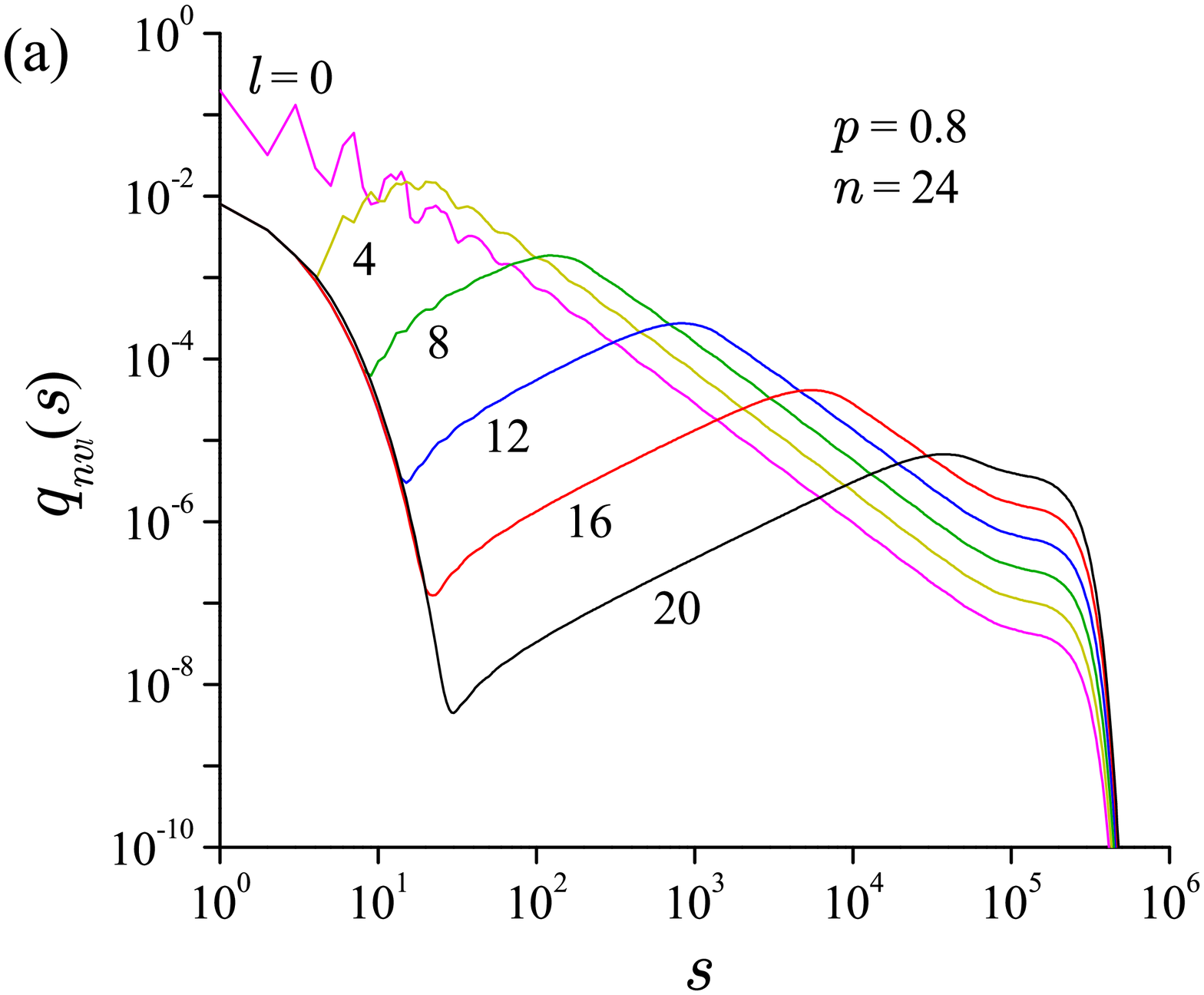}
\includegraphics[trim=20 40 170 10,scale=0.33,clip]{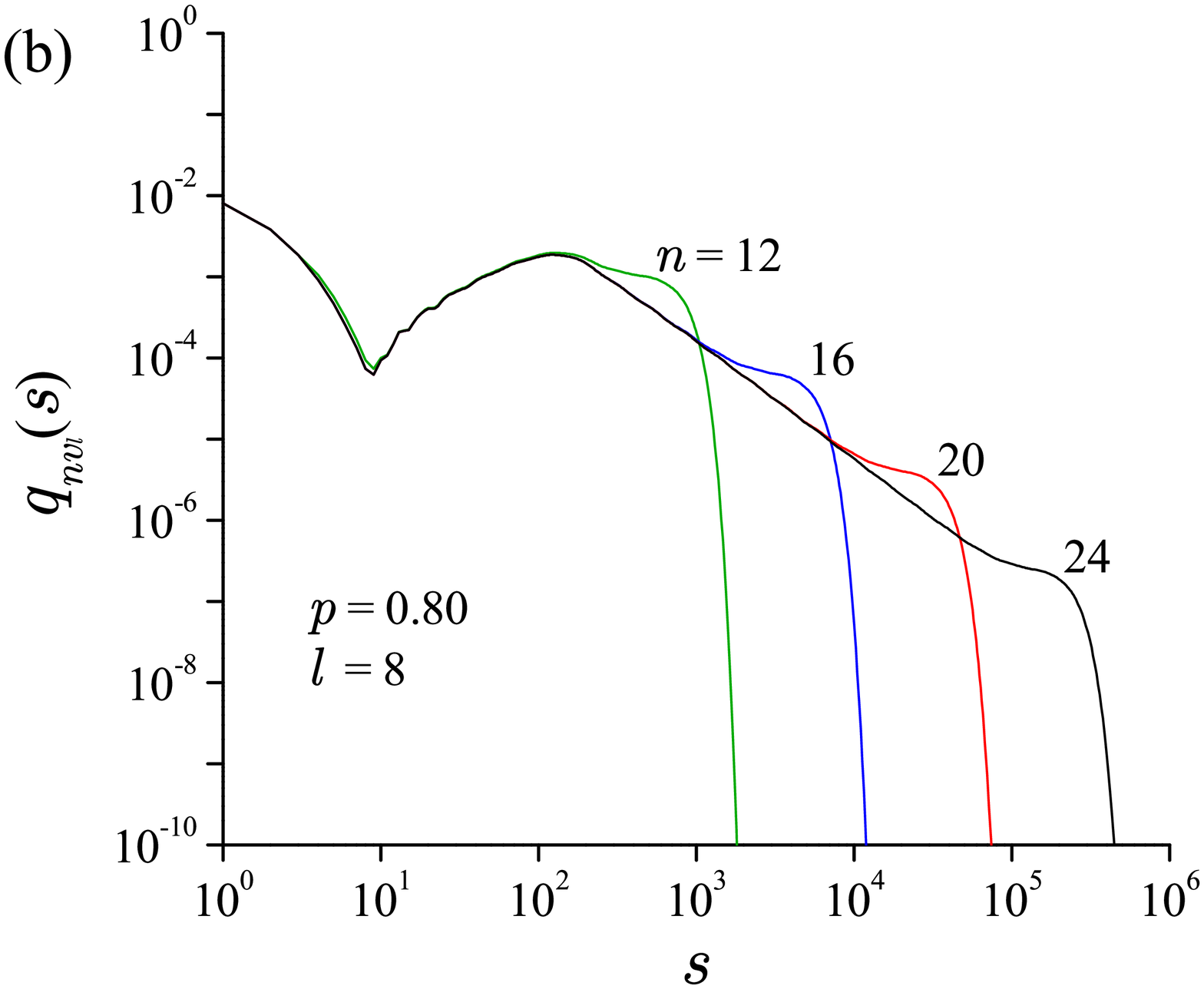}
\\
\includegraphics[trim=20 20 170 10,scale=0.33,clip]{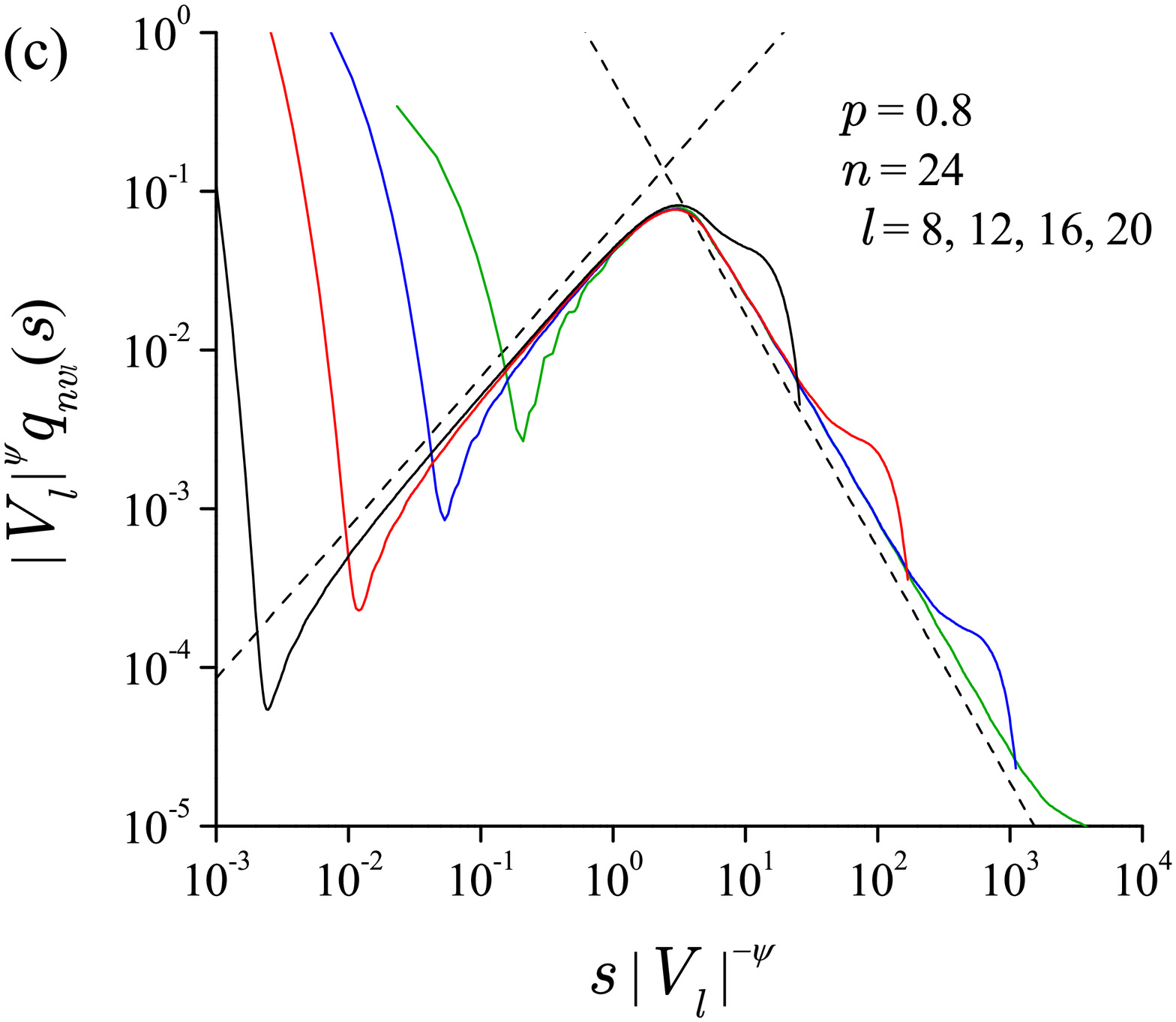}
\includegraphics[trim=20 20 170 10,scale=0.33,clip]{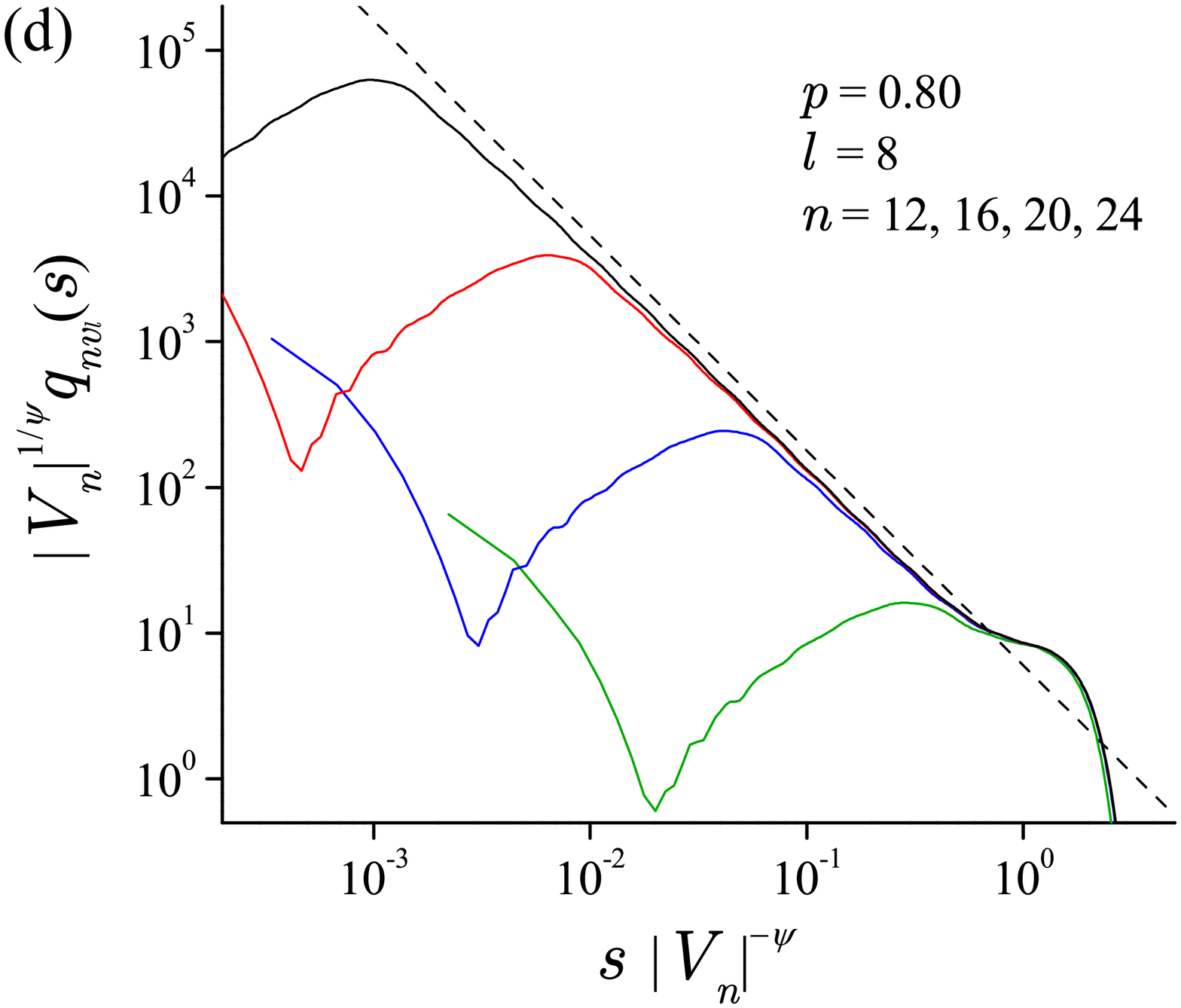}
\end{center}
\vspace{-5mm}
\caption{\label{fig:qnl-s}
(a) $q_{nv_l}(s)$ for various $l$'s.  
(b) $q_{nv_l}(s)$ for various $n$'s. 
(c) Scaling of $q_{nv_l}(s)$ with respect to $|V_l|$. 
The slopes of the dashed lines equal $\tau'$ 
and $-1/\psi$, respectively. 
(d) Scaling of $q_{nv_l}(s)$ with respect to $|V_n|$. The slope of the dashed line equals $-1/\psi$. 
}
\end{figure}
%%%%%%%%%%%%%%%%%%%%%%%%%%%%%%%%%%%%%%%%%%%%%%%%%%%%%%%%%%%%%%%%%

We plot $q_{n v_l}(s)$ for $p=0.8>p_c$ 
with several values of $l$ in Fig.~\ref{fig:qnl-s}(a) 
and of $n$ in Fig.~\ref{fig:qnl-s}(b). 
When $1 \ll |V_l|^\psi \ll |V_n|^\psi$, we observe four regimes; 
\be
q_{nv_l}(s) \left\{
\begin{array}{lcl}
= \qq(s) & \mathrm{for} & s \le l \\
\propto s^{\tau'}   & \mathrm{for} & 1 \ll s \ll |V_l|^\psi \\
\propto s^{-1/\psi} & \mathrm{for} & |V_l|^\psi \ll s \ll |V_n|^\psi \\
\mathrm{rapid \  decay} & \mathrm{for} & s \gg |V_n|^\psi
\end{array}
\right. .
\label{eq:scaling_q_nvl}
\ee
The second crossover from power-law growth to power-law decay occurs 
when $C_{nv_l}$ begins to reach the nearest leaves, 
which are distant from $v_l$ by $l$, probably. 
In this crossover, $q_{nv_l}(s)$ obeys the scaling law  
obtained by replacing $|V_n|$ in Eq.~\eqref{eq:scaling_q_{no}} with $|V_l|$ 
as shown in Fig.~\ref{fig:qnl-s}(c). 
The origin of the power-law decay is roughly understood as follows. 
By letting $q_{nv_l}(j|s)$ be the PDF under the condition that the upper end of $C_{nv_l}$ is $v_{j-1}$ 
and approximating $q_{nv_l}(j|s)$ by $q_{jo}(s)$, we have 
\be
q_{n v_l}(s) &=& \sum_{j=l+1}^{n-1} \pb p^{j-l-1} q_{n v_l}(j|s) + p^{n-l-1} q_{n v_l}(n|s) 
\nonumber \\
& \approx & \frac{\pb}{p^{l+1}} \sum_{j=-\infty}^{\infty} p^{j} q_{jo}(s)
\approx \frac{\pb}{p^{l+1}}  \sum_{j=-\infty}^\infty  |V_j|^{-1} \qqq^> ( s |V_j|^{-\psi} ). 
\label{eq:derive_power-law}
\ee
In the second equation (approximation), 
we assume $|V_l|^\psi \ll s \ll |V_n|^\psi$ 
and modify the lower and upper bounds of the series 
by using that the series is dominated by the terms with $j$ such that $|V_j|^{\psi} \sim s$. 
For the same reason, we ignore the term, $p^{n-l-1} q_{n v_l}(n|s)$. 
In the last equation, 
we ignore the component $q(s)$ in $q_{j+1,o}(s)$
and use the scaling law, Eq.~\eqref{eq:scaling_q_{no}}, 
and $p^j = 2^{(\psi-1)j} \approx |V_j|^{\psi-1}$. 
Equation~\eqref{eq:derive_power-law} satisfies 
$q_{nv_l}(2^{-\psi} s) = 2 q_{nv_l}(s)$, 
which leads to $q_{nv_l}(s) \propto s^{-1/\psi}$. 
More accurate calculation is shown in \ref{sec:leaf_scaling}.

The third crossover around $s \sim |V_n|^\psi$ obeys another scaling law: 
\be
q_{nv_l}(s) = |V_n|^{-1} \qqq_l(s |V_n|^{-\psi} )
= s^{-1/\psi} \tilde{\qqq}_l(s |V_n|^{-\psi} )
\quad \mrm{for} \quad s \gg |V_l|^\psi, 
\label{eq:scaling_q_vl}
\ee
as shown in Fig.~\ref{fig:qnl-s}(d). 
This is derived from Eq.~\eqref{eq:scaling_q_{no}} 
and $\lim_{x \to 0} \tilde{\qqq}_l(x)$ is a finite constant 
(See \ref{sec:leaf_scaling}).

\subsubsection{local fractal exponent}

% Next, we consider $\psi_{v_l}$ defined in Eq.~\eqref{eq:def:vphi}. 
As shown in the Sec.~\ref{sec:pdf4leaf}, 
the large-$s$ behavior is governed by the power-law decay. 
By using Eq.~\eqref{eq:scaling_q_vl}, we have 
\be
P(|C_{nv_l}| > |V_n|^{\vphi}) & \approx & 
% \int_{|V_n|^{\vphi}}^\infty ds |V_n|^{-1} \qqq_l (s |V_n|^{-\psi})
% = |V_n|^{\psi-1} \int_{|V_n|^{\vphi-\psi}}^\infty dx \qqq_l(x). 
\int_{|V_n|^{\vphi}}^\infty ds s^{-1/\psi} \qqqq_l (s |V_n|^{-\psi})
\nonumber \\
& = & |V_n|^{\psi-1} \int_{|V_n|^{\vphi-\psi}}^\infty dx x^{-1/\psi} \qqqq_l(x). 
\ee
% This goes to zero in the limit $n \toinf$ for $\vphi \in (0,1)$. 
For $\vphi > \psi$, both the integral and the prefactor goes to zero. 
For $\vphi \in (0, \psi]$, we have to care about the infrared divergence 
as 
\be
P(|C_{n v_l}|> |V_n|^{\vphi}) 
\approx |V_n|^{\psi-1} |V_n|^{(\vphi- \psi)(1-1/\psi)} \qqqq_l(0)
= |V_n|^{-(1/\psi-1)\vphi} \qqqq_l(0), 
\ee
which vanishes for $n \toinf$ for $\psi<1$. 
Thus, we have 
\be
\psi_{v_l} = 0
\quad \iff \quad 
\lim_{n \toinf} P(|C_{n v_l}| < \infty ) = 1 
\quad \mrm{for} \  p<1 \ \mrm{and} \  l < \infty. 
\ee
The local exponents for vertices near the leaves 
do not coincide with $\psi$ appearing in the scaling equation \eqref{eq:scaling_q_vl}, 
unlike that for vertices near the root.

\subsubsection{moments}

By using Eq.~\eqref{eq:scaling_q_vl}, we have 
\be
\chi_{n v_l}^\mm 
\approx \int_1^\infty ds s^m s^{-1/\psi} \qqqq_l(s |V_n|^{-\psi})
= |V_n|^{\psi^\mm} \int_{|V_n|^{-\psi}}^\infty dx x^{\psi^\mm\!/\psi-1} \qqqq_l(x), 
\label{eq:integral_chi_av_dis}
\ee
where 
$
\psi^\mm \equiv (m+1)\psi - 1. 
$
When 
\be
\psi^\mm > 0 
\ \iff \ 
p > p_c^\mm \equiv 2^{-m/(1+m)}, 
\ee
the integral in the most r.h.s. of Eq.~\eqref{eq:integral_chi_av_dis} 
converges to a finite value for $n\toinf$. 
Thus $\chi_{nv_l}^\mm$ diverges being proportional to $|V_n|^{\psi^\mm}$ for $p>p_c^\mm$. 
For $p<p_c^\mm$, the integral in the infrared region is dominant 
and it converges to 
$\qqqq_l(0) \psi/|\psi^\mm|  \propto (p_c^\mm - p)^{-1}$ 
for $n \toinf$. 
% Note that $\psi^\mm < m \psi$ for $\psi < 1$. 
For $p = p_c^\mm$, $\chi_{nv_l}^\mm$ logarithmically diverges with increasing $n$. 
Thus we have 
\be
\chi_{n v_l}^\mm \propto \left \{
\begin{array}{ccc}
(p_c^\mm - p)^{-1} & \mrm{for} & p \uto p_c^\mm
\\
\ln |V_n| & \mrm{for} & p = p_c^\mm
\\
|V_n|^{\psi^\mm} & \mrm{for} & p > p_c^\mm
\end{array}
\right. 
\ee
for $n \toinf$. 
Note that 
\be
1 > p_c^{(1)} > p_c^{(2)} > \cdots > \lim_{m \toinf} p_c^\mm = p_c. 
\ee
Thus only the infinite-order moment diverges at $p=p_c$. 
This type of weak singularity at $p_c$ is called an infinite-order transition. 
As $p$ increases, the lowest order of the diverging moments decreases.  
% This is in contrast with the fact that 
% $\chi_{n v}^{(m)}$ with $d(o,v)<\infty$ diverges for any $m \ge 1$ at $p_c$. 

%%%%%%%%%%%%%%%%%%%%%%%%%%%%%%%%%%%%%%%%%%%%%%%%%%%%%%%%%%%%
\section{summary and discussions}
\label{sec:summary}

In this paper, we investigated the PDF for a local variable of the CT, namely, 
the size of the cluster including a selected vertex, especially, the origin and a leaf. 
(Although we mainly calculate $q_{nv}(s)$ for the binary tree, 
we remark that $\qq_{nv}(s)$ for the CT is essentially same with $q_{nv}(s)$ 
except unimportant constant factor and the shape of the cutoff of the scaling functions.) 
We found that the PDF for the origin converges to that for BL and that for the leaves converges to a power-law function. 
These behaviors are respectively similar to that in a percolating phase and that at a critical point in Euclidean systems. 
Regarding finite size effects, 
the former has an asymmetric peak with a long tail toward the smaller size direction,
whereas the latter exhibits a simple cutoff. 
These obey their own scaling laws. 
By using the scaling laws, we showed that the local fractal exponent for the origin is positive and that for the leaves is zero. 
These properties correspond to the fact that the probability that 
the vertex belongs to an infinite cluster is positive and zero, respectively. 
In addition, any positive-order moment of the local cluster size for the origin diverges. 
For the leaves, only the infinite-order moment diverges at the critical point 
and the lower-order moments diverge stepwise as $p$ increases. 

% \subsubsection{imhomogeneity}

The power-law of the PDF for the leaves leads to 
the power-law of the global distribution function, that is, the cluster size histogram per vertex;  
\be
n_s \equiv \left\langle \frac{1}{|\VV_n|} 
\sum_{v \in \VV_n} \frac{ \delta_{s |\CC_{nv}|} } {|\CC_{nv}|} 
\right \rangle
= \frac{1}{|\VV_n|} \sum_{v \in \VV_n} \frac{ \qq_{nv}(s) }{s} 
\propto s^{-\tau}. 
\label{eq:ns}
\ee
Since the contribution to the summation is dominated by leaves, 
we have $n_s \approx s^{-1} \qq_{n v_o}(s)$ and therefore $\tau = 1 + 1/\psi$. 
A similar power-law is observed at the critical point 
of an ordinary second-order transition in Euclidean lattices.  
However, we should note that the power-law of Euclidean systems 
is observed also in the local PDF for the bulk region free from the boundary effect 
whereas the power-law directly reflects the boundary effect and the nesting structure of the graph itself in the case of the CT. 
The power-law at a critical point of a finite-dimensional system represents the quasi-long-range coherence, 
i.e., power-law decay of connectedness via open bonds, due to the divergence of a characteristic length. 
In the critical phase of the CT, 
the connectedness between two vertices has short-range coherence decaying exponentially as $p^{d(o,v)}$.

% \subsubsection{order parameter}

Whereas the PDF for the leaves strongly reflects the boundary effect, 
the PDF for the origin is related to that for the vertices in the BL. 
Thus, the observation of the latter enables us to argue the phase of the system 
in the same framework with the infinite homogeneous system, where the phase is well defined. 
When we consider only finite clusters on the BL, 
the PDF is an exponential function as well as in the ordinary percolating phase.
However, the peak due to finite-size effect, $q_n^>(s)$, is significantly different. 
This component can be identified with the PDF of the size of the intersection 
between an infinite cluster on the BL and an $(n-1)$-ball centered at the origin. 
If the supercritical phase was the ordinary percolating phase and there exists a unique infinite cluster,  
the PDF would have a Gaussian peak centered at $s \approx \PP_\infty |\VV_n|$. 
Actually, the PDF has an asymmetric peak whose center position is $O(|\VV_n|^\psi)$ with $\psi \in (0,1)$.

In the percolating phase of a homogeneous system, an infinite cluster is unique and then we have 
\be
\PP_{\infty} \equiv \lim_{\vphi \dto 0} \lim_{n \toinf} P(|\CC_{n \oo}|> |\VV_n|^\vphi) 
= m \equiv \sqrt{ \lim_{n \toinf} \langle |\CC_{n \oo}| \rangle / |\VV_n| }. 
\ee
Here $\PP_\infty$ and $m$ are local quantities. 
This equation does not hold in the critical phase; $\PP_\infty$ is positive but $m$ equals zero 
since $\langle |\CC_{n \oo}| \rangle \propto |\VV_n|^\psi$. 
Note that $\PP_\infty$ is the probability that $\CC_{o}$ in the BL 
is not finite and therefore tells us nothing about the number of infinite clusters. 
We have to distinguish these order parameters. 
We propose to call $m$ an extensive order parameter, 
which is zero in the critical phase and positive in the percolating phase, 
and call $\PP_\infty$ subextensive order parameters, which are positive in both phases. 
We remark that $\PP_\infty > 0 \iff \psi_o > 0$ and $m > 0 \iff \psi_o = 1$.

% \subsubsection{other systems}

We expect all properties described in this paper to apply to percolation on infinite nonamenable graphs (NAGs) 
and the asymptotic limit of the increasing sequence of their subgraph. 
NAGs include BLs, enhanced trees \cite{CCWu97,Nogawa-Hasegawa09} and hyperbolic lattices \cite{Benjamini-Schramm00}. 
It has been proved that there exists a phase where an infinite number of infinite clusters exist 
in the percolation on NAGs \cite{Babson99,Benjamini-Schramm00}. 
% In general, percolation on NAGs exhibits three phases: 
% a nonpercolating phase, a nonuniqueness phase and a percolating (uniqueness) phases. 
In addition, we have preliminarily confirmed that a similar result appears 
in percolation on a hierarchical small-world network, namely, the Faray graph 
\cite{Boettcher12,Nogawa-Hasegawa14}, 
which exhibits a critical phase and a percolating phase. 
Note that the Faray graph is defined as an increasing sequence of a graph 
and cannot be regarded as a subgraph of a well-defined, infinite inhomogeneous graph unlike a CT. 
In the critical phase, the PDF for the vertex with the largest degree is similar to the $\qq_{no}(s)$, 
although the exponent $\tau'$ is negative. 
In addition, the cluster-size histogram is a decaying power function as Eq.~\eqref{eq:ns}. 
Therefore, we expect that it is possible to link the critical phase of the intrinsically inhomogeneous systems 
to the well-defined nonuniqueness phase on infinite homogeneous NAGs, via the property of local variables. 
To confirm the generality of this idea, we need to research the local PDF for various graphs 
where the cluster-size histograms obey the power-law.

\ack

TH is grateful for the financial support from the Grant-in-Aid for Young Scientists (B) (Grant No.~15K17716), 
Grant-in-Aid for Scientific Research (B) (Grant No.~26310203), 
and JST, ERATO, Kawarabayashi Large Graph Project.

%%%%%%%%%%%%%%%%%%%%%%%%%%%%%%%%%%%%%%%%%%%%%%%%%%%%%%%%%%%%
\appendix

\section{Calculation of the PDF for the root vertex}
\label{sec:cal_detail}

\subsection{recursion equation and fixed point}
\label{sec:fixed_point}

By using the fact that $T_{n+1}$ includes two $T_n$'s, 
we obtain a recursion equation: 
\be
Q_{n+1,o}(x) = x [ \pb + p Q_{no}(x) ]^2. 
\label{eq:recursion}
\ee
The fixed {\it point} of this recursion equation, $Q(x)$,  
is given by 
\be
Q(x) = x [ \pb + p Q(x) ]^2 
\quad \iff \quad 
Q(x) = \frac{ 
1 - 2p \pb x \pm \sqrt{ 1 - 4 p \pb x } 
}{ 2 p^2 x }.
\label{eq:fixed_point}
\ee
The stable fixed point for the initial condition $Q_{1o}(x)=x$ is the one denoted by minus sign.
By differentiating Eq.~\eqref{eq:recursion}, 
we obtain the recursion equation of $Q^{(1)}_{no}(x)$ and its fixed point $Q^{(1)}(x)$ as 
\be
Q^{(1)}_{n+1,o}(x) &=& [ \pb + p Q_{no}(x) ] \{ x[ \pb + p Q_{no}(x)] + 2 p Q^{(1)}_{no}(x) \}, 
\\ 
Q^{(1)}(x) &=& \frac{x [ \pb + p Q(x) ]^2}{ 1 - 2 p x [ \pb + p Q(x) ]}. 
\ee

By substituting $Q_{no}(x) = \sums q_{no}(s) x^s$ into Eq.\eqref{eq:recursion}, 
we have 
\be
\sums q_{n+1,o}(s) x^s = 
\pb^2 x + x \sums \left[ 
2 p \pb q_{no}(s)
+ p^2 \sum_{t=1}^{s-1} q_{no}(t) q_{no}(s-t) \right] x^s. 
\ee
By comparing the coefficients for each order of $x^s$, 
we obtain a recursion equation for $s \ge 2$ as 
\be
q_{n+1,o}(s) = 2 p \pb q_{no}(s-1) + p^2 \sum_{t=1}^{s-2} q_{no}(t) q_{no}(s-1-t) 
\label{eq:term_recursion}
\ee
with an initial condition $q_{1o}(s)=\delta_{s1}$ 
and a boundary condition $q_{no}(1)=\pb^2$ for $n \ge 2$.

For $\TT_n$, we have the recursion equations and the fixed point as 
\be
\QQ_{n+1,o}(x) &=& x [ \pb + p Q_{no}(x) ]^3 = Q_{n+1,o}(x) [ \pb + p Q_{no}(x) ], 
\label{eq:recursion_G}
\\
\qq_{n+1,o}(s) &=& \pb q_{n+1,o}(s) + p \sum_{t=1}^{s-1} q_{n+1,o}(t) q_{no}(s-t), 
\label{eq:recursion_g}
\\
\QQ(x) &=& Q(x) [ \pb + p Q(x) ], 
\\ 
\qq(s) &=& \pb q(s) + p \sum_{t=1}^{s-1} q(t) q(s-t). 
\label{eq:rec_q_infty}
\ee

\subsection{function form of $q(s)$}
\label{sec:series_expansion}

By using an equality 
\be
\sqrt{1-y} = 1 - \frac{y}{2} - 
\sum_{n=2}^\infty \frac{(2n-3)!}{n! (n-2)!} \frac{y^n}{2^{2n-2}} 
\ee
in Eq.~\eqref{eq:fixed_point}, we have 
% \be
% Q(x) = 2\, \frac{\pb}{p} \sum_{n=1}^\infty 
% \frac{(2n-1)!}{(n+1)! (n-1)!} (p\pb x)^n,  
% \ee
% and 
\be
q(s) = 2\, \frac{\pb}{p} \frac{(2s-1)!}{(s+1)! (s-1)!} (p\pb)^s. 
\label{eq:q_infty}
\ee
We remark that $p^2 q(s)$ is a function of $p \pb$ 
and is invariant against the replacement of $p$ with $\pb$, 
i.e., the reflection at $p=1/2$. 
By multiplying the both sides of Eq.~\eqref{eq:rec_q_infty} by $p^3$, 
we can show that $p^3 \qq(s)$ is also a function of $p \pb$.

Equation~\eqref{eq:q_infty} leads to 
\be
&&
\frac{q(s+1)}{q(s)} = \frac{s+1/2}{s+2} (4p \pb). 
\ee
For $p=1/2$, 
we have  
$q(s+1) - q(s) = -[3/2(s+2)]q(s)$. 
For $s \gg 1$, we have 
\be
\frac{d q(s)}{d s} \approx -\frac{3}{2} \frac{q(s)}{s} 
\ \iff \ 
q(s) \propto s^{-3/2}.
\ee
For arbitrary $p$ and $s \gg 1$, we have 
\be
q(s) = (4p\pb)^s q(s)\Big{|}_{p=p_c} \propto s^{-3/2} (4p\pb)^s. 
\ee

%%%%%%%%%%%%%%%%%%%%%%%%%%%%%%%%
\subsection{evaluation of $\tau'$}
\label{sec:tau'}

% Here we construct the recursion equation for $q_{no}^>(s)$ 
% to evaluate $\tau'$. 

The deviation of the GF from the fixed point,
\be
Q_{no}(x) - Q(x) \equiv Q_{no}^>(x) = \sums q_{no}^>(s) x^s, 
% q_{no}^>(s) \equiv q_{no}(s) - q(s),
\label{eq:R>}
\ee
satisfies 
\be
Q_{n+1,o}^>(x) &=& px \left\{ 2 \pb + p [ 2 Q(x) + Q_{no}^>(x)] \right\} Q_{no}^>(x),  
\label{eq:rec_R>}
\\
q_{n+1,o}^> \left(s \right) &=& 2 p \pb q_{no}^> ( s-1 )
+ 2 p^2 \sum_{t=1}^{s-2} q(t) q_{no}^>( s-1-t ) 
\nonumber \\
& & + p^2 \sum_{t=1}^{s-2} q_{no}^>(t) q_{no}^> ( s-1-t ). 
\label{eq:rec_q>}
\ee
By substituting $q_{no}^>(s) = |V_n|^{-\psi}  \qqq^>(s |V_n|^{-\psi})$ 
to Eq.~\eqref{eq:rec_q>}, we obtain
\be
\qqq^> \left(s h^{-(n+1)} \right) 
&=& 2 p \pb h \qqq^> \left( (s-1) h^{-n} \right)
+ 2 p^2 h \sum_{t=1}^{s-2} q(t) \qqq^> \left( (s-1-t) h^{-n} \right)
\nonumber \\
& & + p^2 h^{1-n} \sum_{t=1}^{s-2} \qqq^>( t h^{-n} ) \qqq^> \left( (s-1-t) h^{-n} \right). 
\ee
Here we put $h \equiv 2^\psi (= 2p)$ and use $|V_n|^\psi \approx h^n$. 
The third term on the r.h.s., which is less than 
$p^2 h^{1-n} \sum_{t=1}^{s} \qqq^>(s h^{-n} )^2 
\propto p^2 h (s h^{-n} )^{1 + 2\tau'}$, 
is negligible for $s \ll h^n$. 
By putting $s h^{-n}=x$, we have 
\be
\qqq^>(x/h) &\approx& 2 p \pb h \qqq^>(x) 
+ 2 p^2 h  \sumt q(t) \qqq^> \left(x - (1+t) h^{-n} \right)
\nonumber \\
&\approx & 2 p h [ \pb + p Q(1)] \qqq^>(x) 
= 4p \pb \qqq^>(x). 
\ee
In the second approximation, we assume $x \gg h^{-n}$. 
Finally, we obtain
\be
h^{-\tau'} = \frac{\qqq^>(x/h)}{\qqq^>(x)} = 4 p \pb 
\ \iff \ 
\tau' = - \frac{\ln 2 \pb}{\ln 2 p} - 1.
\ee

%%%%%%%%%%%%%%%%%%%%%%%%%%%%%%%%
\section{scaling of $q_{nv}$ near the leaves}
\label{sec:leaf_scaling}

Let $v_l$ be a vertex in the $(n-l-1)$th layer of $T_n$ with $n \ge l+1$.
By defining 
\be
A_{nl}(x) \equiv \sums a_{nl} x^s, \quad
a_{nl}(s) \equiv P\left( |C_{n v_l}|=s \land  o \in C_{n v_l} \right), 
\\
B_{nl}(x) \equiv \sums b_{nl} x^s, \quad
b_{nl}(s) \equiv P\left( |C_{n v_l}|=s \land  o \notin C_{n v_l} \right), 
\ee
we have recursion equations 
\be
Q_{nv_l}(x) &=& A_{nl}(x) + B_{nl}(x), 
\\
A_{n+1,l}(x) &=& p x [ \pb + p Q_{no}(x) ] A_{nl}(x), 
\label{eq:rec_A}
\\
B_{n+1,l}(x) &=& B_{nl}(x) + \pb A_{nl}(x). 
\label{eq:rec_B}
\ee
The initial condition for $n=l+1$ is given by 
$A_{l+1,l} (x) = Q_{lo}(x)$ and $B_{l+1,l}(x)=0$ 
because $v_l$ is the root of $T_{l+1}$.
We have $B_{nl}(x) = \pb \sum_{j=l+1}^{n-1} A_{nl}(x)$.

Here we solve Eq.~\eqref{eq:rec_A} by using Eq.~\eqref{eq:rec_R>}. 
By letting $x Q_{no}^>(x) \equiv S_n(x)$ in Eq.~\eqref{eq:rec_R>}, we have
\be
S_{n+1}(x) &=& p \left\{  2 x [ \pb + p Q(x) ] + p S_n(x) \right \} S_n(x), 
\\
S_{n+1}'(x) &=& S_n(x) \frac{d}{dx} 2px [ \pb + p Q(x) ] 
\nonumber \\
& & + 2 p x \left[ \pb + p Q(x) + \frac{p}{x} S_n(x) \right] S_n'(x).
\label{eq:rec_S_app}
\ee
For $p>p_c$ and $x \ge 1$, $S_n(x)$ and  $S'_n(x)$ diverge for $n \toinf$ 
while $Q(x)$ and $Q'(x)$ remain finite. 
This allows us to ignore the first term in Eq.~\eqref{eq:rec_S_app} 
and then we have 
% From Eq.~\eqref{eq:rec_A} and , we obtain 
\be
S_{n+1}'(x) \approx 2 p x [ \pb + p Q_{no}(x) ] S_n'(x) 
\ \To \ 
\frac{A_{n+1,l}(x)}{S_{n+1}'(x)} \approx \frac{1}{2}\frac{A_{nl}(x)}{S_{n}'(x)}
\ee
for $l \ll n$. 
Thus, we have 
\be
A_{nl}(x) & \propto & 2^{-n} S_n'(x) 
\propto 2^{-n} \frac{d}{dx} x Q_{no}^>(x), 
\\
a_{nl}(s) &\propto& 2^{-n} (s+1) q_{no}^>(s) 
\approx 2^{-n} s |V_n|^{-\psi} \qqq^>( s |V_n|^{-\psi} ) 
% \nonumber
% \ee and  \be a_{nl}(s) 
\nonumber \\
&=& |V_n|^{-1} \tilde{a}_l( s |V_n|^{-\psi} )
= s^{-1/\psi} \tilde{\tilde{a}}_l( s |V_n|^{-\psi} ), 
\label{eq:scaling_a_nl}
\ee
where $\tilde{a}_l(x) \propto x \qqq^>(x) \propto x^{1+\tau'}$ and 
$\tilde{\tilde{a}}_l(x) \equiv x^{1+1/\psi} \qqq^>(x)$.

Next, we show that $b_{nl}(s)$ obeys a similar scaling law by an inductive method. 
If we assume that 
\be
s^{1/\psi} b_{nl}(s) 
= \pb \sum_{j=l+1}^{n-1} \tilde{\tilde{a}}_{l} (s |V_j|^{-\psi} ) 
\equiv \tilde{\tilde{b}}_l( s |V_n|^{-\psi} ), 
\ee
we have 
\be
s^{1/\psi} b_{n+1,l}(s) 
&=& \pb \sum_{j=l+1}^{n} \tilde{\tilde{a}}_{l} (s |V_j|^{-\psi}) 
= \pb \sum_{j=l}^{n-1} \tilde{\tilde{a}}_{l} (s |V_{j+1}|^{-\psi}) 
\nonumber \\
&=& \tilde{\tilde{b}}_l( s |V_{n+1}|^{-\psi} ) 
+ \pb \tilde{\tilde{a}}_l( s |V_{l+1}|^{-\psi} ). 
\ee
The second term on the most r.h.s. is negligible for $s \gg |V_{l+1}|^\psi$.  
Furthermore, these lead to the conclusion that $q_{n v_l}(s) = a_{nl}(s) + b_{nl}(s)$ 
obeys the scaling law, as Eq.~\eqref{eq:scaling_a_nl}.

Finally, we show that 
\be
\lim_{x \to 0} \tilde{\tilde{q}}_l(x)
= \lim_{x \to 0} \tilde{\tilde{b}}_l(x) 
\in (0, \infty).
\label{eq:tilde_b(0)}
\ee
By putting $s = |V_k|^\psi \approx h^k$ with $h \equiv 2^\psi$, we have 
\be
(\pb)^{-1} b_{nl}(h^k)
&=& \sum_{j=l+1}^{n-1} a_{nl}(h^k)
= \sum_{j=l+1}^{n-1} h^{-j/\psi} \tilde{a}_l( h^{k-j} )
% \nonumber \\
% &\propto& \sum_{j=-\infty}^{\infty} h^{-j/\psi} h^{(k-j)(1+\tau')} 
% &=& \tilde{a}_l(0)  \sum_{j'=0}^{n-1-k} h^{-(j'+k)/\psi - j'(1+\tau')}
\nonumber \\
&=& h^{-k/\psi} \sum_{j'=l-k+1}^{n-k+1} h^{-j'/\psi} \tilde{a}(h^{-j'}).
\ee
We use Eq.~\eqref{eq:scaling_a_nl} in the second equation. 
For $h^l \ll h^k \ll h^n$, 
the summation in the most r.h.s. is independent of $k$. 
Thus, we have $b_{nl}(s) \propto s^{-1/\psi}$ for $|V_l|^\psi \ll s \ll |V_n|^\psi$ 
and Eq.~\eqref{eq:tilde_b(0)} holds.

% \newpage \input{cluster_density.tex} \newpage

\section*{References}

% \bibliography{d:/nogawa/study/manuscripts/mybib.bib}
% \input{BinaryTree160316.bbl}
\input{BinaryTree.bbl}

\end{document}

%% file: BinaryTree.bbl
\providecommand{\newblock}{}